# Cryogenic spectroscopy of ultra-low density colloidal lead chalcogenide quantum dots on chip-scale optical cavities towards single quantum dot near-infrared cavity QED

Ranojoy Bose*, Jie Gao, James F. McMillan, Alex D. Williams and Chee Wei Wong*

*Optical Nanostructures Laboratory, Center for Integrated Science and Engineering, Solid-state Science and Engineering, and Mechanical Engineering, Columbia University, New York, NY 10027*
*rb2261@columbia.edu, cww2104@columbia.edu*

**Abstract:** We present evidence of cavity quantum electrodynamics from a sparse density of strongly quantum-confined Pb-chalcogenide nanocrystals (between 1 and 10) approaching single-dot levels on moderately high-*Q* mesoscopic silicon optical cavities. Operating at important near-infrared (1500-nm) wavelengths, large enhancements are observed from devices and strong modifications of the QD emission are achieved. Saturation spectroscopy of coupled QDs is observed at 77K, highlighting the modified nanocrystal dynamics for quantum information processing.

## References and links

1. H. Mabuchi and A. C. Doherty, "Cavity Quantum Electrodynamics: Coherence in Context", Science **298,** 1372 (2002)
2. G. Khitrova, H. M. Gibbs, M. Kira, S. W. Koch, and A. Scherer, "Vacuum Rabi splitting in semiconductors," Nature Physics **2**, 81 (2006)
3. K. Srinivasan and O. Painter, "Linear and nonlinear optical spectroscopy of a strongly coupled microdisk-quantum dot system," Nature **450**, 862 (2007)
4. Y.-F. Xiao, J. Gao, X.-B. Zou, J. F. McMillan, X. Yang, Y.-L. Chen, Z.-F. Han, G.-C. Guo, and C. W. Wong, "Coupled quantum electrodynamics in photonic crystal cavities towards controlled phase gate operations," New. J. Phys. **10**, 123013 (2008)
5. M. Pelton, C. Santori, J. Vučković, B. Zhang, G. S. Solomon, J. Plant and Y. Yamamoto, "Efficient Source of Single Photons: A Single Quantum Dot in a Micropost Microcavity," Phys. Rev. Lett. **89**, 233602 (2002)
6. S. Strauf, N. G. Stoltz, M. T. Rakher, L. A. Coldren, P. M. Petroff, and D. Bouwmeester, "High-frequency single-photon source with polarization control," Nature Photonics **1**, 704 (2007)
7. C. Santori, D. Fattal, J. Vučković, G. S. Solomon, Y. Yamamoto, "Indistinguishable photons from a single-photon device," Nature **419**, 594 (2002)
8. F. W. Sun and C. W. Wong, "Indistinguishability of independent single photons," Phys. Rev. A **79**, 013824 (2009)
9. A. Faraon, I. Fushman, D. Englund, N. Stoltz, P. Petroff, and J. Vučković, "Coherent generation of nonclassical light on a chip via photon-induced tunneling and blockade," Nature Physics **4**, 859 (2008)
10. K. Hennessy, A. Badolato, M. Winger, D. Gerace, M. Atatüre, S. Gulde, S. Fält, E. Hu, and A. Imamoğlu, "Quantum nature of a strongly coupled single quantum dot-cavity system," Nature **445**, 896 (2007)
11. J. P. Reithmaier, G. Sęk, A. Löffler, C. Hofmann, S. Kuhn, S. Reitzenstein, L. V. Keldysh, V. D. Kulakovskii, T. L. Reinecke, A. Forchel, "Strong coupling in a single quantum dot-semiconductor microcavity system," Nature **432**, 197 (2004)
12. M. V. Gurudev Dutt, L. Childress, L. Jiang, E. Togan, J. Maze, F. Jelezko, A. S. Zibrov, P. R. Hemmer, and M. D. Lukin, "Quantum register based on individual electronic and nuclear spin qubits in diamond," Science **316**, 1312 (2007)
13. Y. Shen, T. M. Sweeney, and H. Wang, "Zero-phonon linewidth of single nitrogen vacancy centers in diamond nanocrystals," Phys. Rev. B **77**, 033201 (2008)

14. S. Kako, C. Santori, K. Hoshino, S. Gotzinger, Y. Yamamoto, Y. Arakawa, "A gallium nitride single-photon source operating at 200K," Nature Mat. **5**, 887 (2006)
15. C. B. Poitras, M. Lipson, M. A. Hahn, H. Du, and T. D. Krauss, "Photoluminescence enhancement of colloidal semiconductor quantum dots embedded in a monolithic microcavity," Appl. Phys. Lett. **82**, 4032 (2003)
16. I. Fushman, D. Englund, and J. Vučković, "Coupling of PbS quantum dots to photonic crystal cavities at room temperature," Appl. Phys. Lett. **87**, 241102 (2005)
17. R. Bose, X. Yang, R. Chatterjee, J. Gao, C.W. Wong, "Weak coupling interactions of colloidal lead sulphide nanocrystals with silicon photonic crystal nanocavities near 1.55 μm at room temperature," Appl. Phys. Lett. **90**, 111117 (2007)
18. Z. Wu, Z. Mi, P. Bhattacharya, T. Zhu, J. Xu, "Enhanced spontaneous emission at 1.55 μm from colloidal PbSe quantum dots in a Si photonic crystal microcavity," Appl. Phys. Lett. **90**, 171105 (2007)
19. R. Bose, D. V. Talapin, X. Yang, R. J. Harniman, P. T. Nguyen, and C. W. Wong, "Interaction of infilitrated colloidal PbS nanocrystals with high Q/V silicon photonic bandgap nanocavities for near-infrared enhanced spontaneous emissions," Proc. SPIE **6005**, 600509 (2005)
20. A. G. Pattantyus-Abraham, H. Qiao, J. Shan, K. A. Abel, T-S Wang, F. C. J. M. van Veggel and J. F. Young, "Site-Selective Optical Coupling of PbSe Nanocrystals to Si-Based Photonic Crystal Microcavities", *Nano Lett.* (to be published)
21. S. Vignolini, F. Riboli, F. Intonti, M. Belotti, M. Gurioli, Y. Chen, M. Colocci, L. Claudio Andreani, and D. S. Wiersma, "Local nanofluidic light sources in silicon photonic crystal microcavities,", Phys. Rev. E. **78**, 045603 (2008)
22. A. I. Akimov, Al. L. Efros, A. A. Onushchenko, "Quantum size effect in semiconductor nanocrystals," Solid State Comm. **56**, 921 (1985)
23. L. E. Brus, "Electron-electron and electron-hole interactions in small semiconductor crystallites: The size dependence of the lowest excited excitonic state," J. Chem. Phys. **80**, 4403 (1984)
24. F. W. Wise, "Lead Salt quantum dots: The limit of strong quantum confinement," Acc. Chem. Res. **33**, 773 (2000)
25. J. Warner, E. Thomsen, A. R. Watt, N. R. Heckenberg, H. Rubinsztein-Dunlop, "Time-resolved photoluminescence spectroscopy of ligand-capped PbS nanocrystals," Nanotech. **16**, 175 (2005)
26. S. W. Clark, J. M. Harbold, F. W. Wise, "Resonant energy transfer in PbS quantum dots," J. Phys. Chem. C **111**, 7302 (2007)
27. L. Cademartiri, J. Bertolotti, R. Sapienza, D. S. Wiersma, G. von Freymann, G. A. Ozin, " Multigram Scale, Solventless, and Diffusion-Controlled Route to Highly Monodisperse PbS Nanocrystals," J. Phys. Chem. B. **110**, 671 (2006)
28. R. Bose, R, J. F. McMillan, J. Gao, C. J. Chen, D. V. Talapin, C. B. Murray, K. M. Rickey, and C. W. Wong, "Temperature-tuning of near-infrared monodisperse quantum dots at 1.5 μm for controllable Förster energy transfer," Nano Lett. **8**, 2006 (2008)
29. E. M. Purcell, "Spontaneous emission probabilities at radio frequencies," Phys. Rev. **69**, 681 (1946)
30. T. Tanabe, M. Notomi, E. Kuramochi, A. Shinya, H. Taniyama, "Trapping and delaying photons for one nanosecond in an ultrasmall high-$Q$ photonic-crystal nanocavity," Nature Photonics **1**, 49 (2006)
31. S. Noda, M. Fujita, and T. Asano, "Spontaneous-emission control by photonic crystals and nanocavities," Nature Photonics **1**, 449 (2007)
32. A. Farjadpour, D. Roundy, A. Rodriguez, M. Ibanescu, P. Bermel, J. D. Joannopoulos, S. G. Johnson, and G. Burr, "Improving accuracy by subpixel smoothing in the finite-difference time-domain," Opt. Lett. **31**, 2972 (2006)
33. S. Kocaman, R. Chatterjee, N. C. Panoiu, J. F. McMillan, M. B.Yu, R. M. Osgood, D. L. Kwong, and C. W. Wong, "Observations of zero-order bandgaps in negative-index photonic crystal superlattices at the near-infrared," Phys. Rev. Lett. **102**, 203905 (2009)
34. M. W. McCutcheon, G. W. Rieger, I. W. Cheung, J. F. Young, D. Dalacu, S. Frederick, P. J. Poole, G. C. Aers, and R. L. Williams, "Resonant scattering and second-harmonic spectroscopy of planar photonic crystal nanocavities," Appl. Phys. Lett. **87**, 221110 (2005)
35. P. B. Deotare, M. W. McCutcheon, I. W. Frank, M. Khan, and M. Loncar, "High quality factor photonic crystal nanobeam cavities," Appl. Phys. Lett. **94**, 121106 (2009)
36. C. B. Murray, S. Sun, W. Gaschler, H. Doyle, T. A. Betley, C. R. Kagan, "Colloidal synthesis of nanocrystals and nanocrystal superlattices," IBM J. Res. &. Dev. **45**, 47 (2001)
37. M. A. Hines, G. D. Scholes, "Colloidal PbS nanocrystals with size-tunable near-infrared emission: observation of post-synthesis self-narrowing of the particle size distribution," Adv. Mater. **15**, 1844 (2003)
38. D. V. Talapin and C. B. Murray, "PbSe nanocrystal solids for n- and p-channel thin film field-effect transistors," Science **310**, 86 (2005)
39. J. M. Pietryga, K. K. Zhuravlev, M. Whitehead, V. I. Klimov, and R. D. Schaller, "Evidence of barrierless Auger recombination in PbSe nanocrystals: A pressure-dependent study of transient optical absorption," Phys. Rev. Lett. **101**, 217401 (2008)
40. J. C. Johnson, K. A. Gerth, Q. Song, J. E. Murphy, and A. J. Nozik, and G. D. Scholes, "Ultra-fast exciton fine structure relaxation dynamics in lead-chalcogenide nanocrystals," Nano Lett. **8**, 1374 (2008)

41. P. Michler, A. Imamoğlu, M. D. Mason, P. J. Carson, G. F. Strouse and S. K. Buratto, "Quantum correlation among photons from a single quantum dot at room temperature," Nature **406**, 968 (2000)
42. X. Brokmann, G. Messin, P. Desbiolles, E. Giaocobino, M. Dahan, J. P. Hermier, "Colloidal CdSe/ZnS quantum dots as single-photon sources," New. J. Phys. **6**, 99 (2004)
43. N. Le Thomas, U. Woggon, O. Schöps, M. V. Artemyev, M. Kazes, and U. Banin, "Cavity QED with semiconductor nanocrystals," Nano Lett. **6**, 557 (2006)
44. G. Allan, C. Delerue, "Confinement effects in PbSe quantum wells and nanocrystals," Phys. Rev. B **70**, 245321 (2004)
45. J. M. An, A. Franceschetti, and A. Zunger, "The excitonic exchange splitting and radiative lifetime in PbSe quantum dots," Nano Lett. **7**, 2129 (2007)
46. J. J. Peterson and T. D. Krauss, "Fluorescence spectroscopy of single lead sulfide quantum dots," Nano Lett. **6**, 510 (2006)
47. L. Cademartiri, E. Montanari, G. Calestani, A. Migliori, A. Guagliardi, G. A. Ozin, "Size Dependent Extinction Coefficients of PbS Quantum Dots," J. Am. Chem. Soc. **128**, 10337 (2006)
48. A. Badolato, K.Hennessy, M. Atatüre, J. Dreiser, E. Hu, P. M. Petroff, and A. Imamoğlu, "Deterministic coupling of single quantum dots to single nanocavity modes," Science **308**, 1158 (2005)
49. L. Turyanska, A. Patane, M. Henini, B. Hennequin, N. R. Thomas, "Temperature dependence of the photoluminescence emission from thiol-capped PbS quantum dots," Appl. Phys. Lett. **90**, 101913 (2007)
50. A. Olkhovets, R. -C. Hsu, A. Lipovskii, F. W. Wise, "Size-dependent variation of the energy gap in lead-salt quantum dots," Phys. Rev. Lett. **81**, 3539 (1998)
51. G. T. Reed and A. P. Knights, *Silicon Photonics: An introduction* (Wiley, 2004)
52. A. Auffèves, J.-M. Gérard, and J.-P. Poizat, "Pure emitter dephasing: a resource for advanced solid-state single-photon sources," Phys. Rev. A **79**, 053838 (2009)
53. I. Kang and F. W. Wise, "Electron structure and optical properties of PbSe quantum dots," J. Opt. Soc. Am. B **14**, 1632 (1997)
54. G. Allan, C. Delerue, "Confinement effects in PbSe quantum wells and nanocrystals," Phys. Rev. B **70**, 245321 (2004)
55. J. M. Gérard, B. Gayral, "Strong Purcell effect for InAs quantum boxes in three-dimensional solid-state microcavities," J. Lightwave Tech. **17**, 2089 (1999)
56. B. Mahler, P. Spinicelli, S. Buil, X. Quelin, J.-P. Hermier, and B. Dubertret, "Towards non-blinking colloidal quantum dots," Nature Mat. **7**, 659 (2008)
57. V. Fomenko and D. J. Nesbitt, "Solution control of radiative and nonradiative lifetimes: a novel contribution to quantum dot blinking suppression," Nano Lett. **8**, 287 (2008).

**1. Introduction**

Cavity quantum electrodynamics (cQED) experiments have been strongly motivated by the need for an efficient on-demand single-photon source that is relevant in many quantum cryptography and quantum information processing applications [1-4]. Charged carriers, or excitons, in single quantum dots (QD) can be excited in a controllable way to inhibit multi-photon emission, enabling a sub-Poissonian single-photon source [5,6]. Coupling of QDs to photonic structures result in a faster recombination rate for the excitons—overcoming problems due to decoherence in single photon source applications [7,8].

As an alternative to the highly mature domain of self-assembled QDs [9-11] to study cQED in photonic structures, several novel systems have been proposed—such as nitrogen-vacancy centers in diamond [12,13] and strongly quantum-confined nanomaterials such as III-nitrides binary materials with large optical phonon energies to operate at 200K temperatures [14]. Nanocrystals formed through synthetic routes, such as CdSe and PbS nanocrystals, have also been used in successfully demonstrating coupled QD-cavity interactions in photonic structures [15,16], and offer important possibilities towards scalable quantum computation that are presently unprecedented in self-assembled technologies. Particularly promising candidates for QED applications are the Pb-chacolgenide nanocrystals, that can be post-integrated with the vast silicon processing infrastructure [17, 18], spatially positioned through electron-beam lithography in a host resist matrix [16,17,19] or other novel techniques [20,21], and have exciton ground state transitions in the near-infrared (1.55-µm) for direct compatibility with the embedded fiber communications network. PbS (Se) QDs also exhibit large exciton Bohr radii compared to the physical dot sizes, resulting in strong quantum confinement [22-24], and devices incorporating these QDs are capable of room temperature operation [16-21] allowing remarkable possibilities for integrated photon sources. Prior work in silicon cavities has been limited to large ensembles of QDs (>10,000) [16-21] due to poorer detection efficiency of near-infrared photon detectors as well as the long lifetimes of the PbS QDs, reported to be around 1-2 μs in solution [25-28] and between 100 ns (at 1500 nm) [27] and 2 μs (at 900 nm) [26] in films. Here we examine a few Pb-chalcogenide QDs coupled to moderately high-$Q$ heterostructured photonic crystal cavities approaching the single quantum dot limit, enabled through coupling interactions of the QDs with the cavities. Through non-resonant photoluminescence spectroscopy, excited state saturation is observed at 77 K as an effort to determine the Purcell factor [29], with large (15×) emission enhancements observed for the few QDs on-resonance with the cavity mode.

## 2. Cavity system

Our mesoscopic optical cavity (Figure 1a) consists of a silicon-on-insulator heterostructured photonic crystal lattice [30,31], which confines modes exhibiting wavelength-scale [$\sim 1.2(\lambda/n)^3$] volumes. Our design has heterostructured lattices $a_1$, $a_2$, and $a_3$ of 410, 415 and 420 nm respectively to achieve mode-gap-type confinement. These designed optical cavities offer a smoother electric field envelope function at the heterostructure boundaries for higher $Q$s, differing from earlier QD-cavity studies involving point-defect cavities. The optical cavity has designed air hole radii $r$ of (118-, 124-, and 130-nm) with a thickness $t$ of 0.61 $a_1$ (250 nm) on a $SiO_2$ insulator substrate. The designed cavity field profiles ($E_x$ which are the dominant modes) (Figure 1 c) are calculated from complete 3D finite-difference time-domain (FDTD) that solves the time-dependent Maxwell's equations with subpixel accuracy [32]. The side profiles of the fields emphasize that while the surface QDs do not see the field maximum, they are still able to couple effectively into the mode due to the evanescent field at the silicon/air interface, with 44% of the field amplitude at the silicon/air interface instead of the cavity maxima. The sparse density of surface QDs (Figure 1a) does not significantly change the cavity mode field distribution.

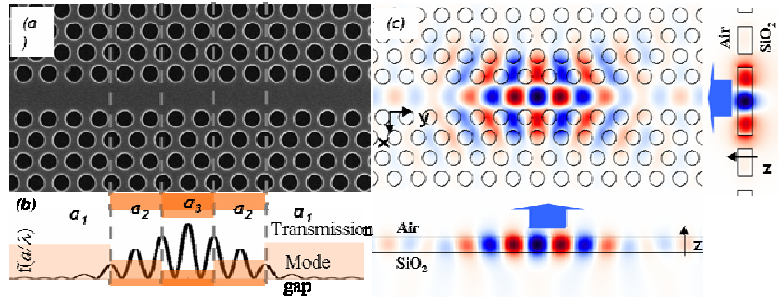

Fig. 1 (a) Scanning electron micrograph (SEM) of a multistep heterostructure cavity with few quantum dots at the cavity, on the device surface. (b) Schematic of the mode-gap cavity confinement. The white regions show photonic bandgaps for in-plane wavevectors. The curve represents 1-D field intensity ($|E_x|^2$) computed using 3D FDTD (c) Top- and side views of the cavity field-mode ($|E_x|^2$) computed with the 3D FDTD method. $r=.3024a_1$, where the other definitions are the same as in figure $a$.

## 3. Device Fabrication and Experiments

Our samples are fabricated in a 248 nm lithography CMOS foundry, with low (sub-20 Å) statistically quantified disorder [33]. This fabrication technique allows for on-chip devices with near-identical device performances and quality factors for the same design, and validates the mature silicon CMOS infrastructure for fabricating near-infrared optical components. The cavities consist of two primary designs (6- or 40-linearly missing holes), with varying waveguide-cavity coupling for planar characterizations. We use resonant cross-polarization spectroscopy [34,35] for characterization of the cavity mode and find this technique to be more reliable than traditional waveguide based measurements for passively characterizing the system, due to the inherent device architecture. The cavities typically exhibit $Q$s between 200 and 500, although our 3D FDTD calculations predict higher $Q$s (theoretical $Q$ 5240). The discrepancy is primarily due to fabrication for a cavity mode that admits low tolerance in error. The colloidal Pb-chalcogenide (PbS) QDs are synthesized using standard methods [36-38], dispersed in chloroform, and exhibit a photoluminescence spectrum centered at 1460 nm with a large spectral width $\Delta\lambda$ of ~ 150 nm at room temperature with a <10% size dispersion. The QDs are obtained from Evident Technologies, and are carefully integrated through a spin-coating procedure that results in a sparse distribution of single dots on the device surface (Figure 2a). Through passive cross-polarization characterization measurements of the cavities with and without quantum dots, we can confirm that the surface QD do not induce any degradation in the cavity $Q$.

PbS QDs are reported to possess a barrierless rapid Auger recombination that is controlled by the QD size [39,40], allowing multi-exciton suppression and antibunching for single-photon applications [41,42]. The Pb-chalcogenide system also has order-of-magnitude comparable oscillator strengths to CdSe colloidal QDs, supporting efforts in strong light-matter coupling [43-45]. At sub-1-um wavelengths with silicon detectors, recent measurements have remarkably isolated single Pb-chalcogenide QDs [46], although the linewidths are unusually large, which may be a result of spectral diffusion and power broadening. At the longer 1.55 µm near-infrared communication wavelengths, detection of a few or a single Pb-chalcogenide QD is even more challenging, even with our state-of-the-art non-silicon detectors and avalanche photodiodes, and must rely on cavity-enhanced spontaneous emission dynamics to shorten the radiative lifetime for enhanced photon counts. As we show below, our cavity-enhanced experiments allow for the detection of approximately less than 10 QDs at 1.55 µm wavelengths.

We are able to visually image single PbS QDs using a scanning electron microscope (SEM) and confirm it with atomic force microscopy measurements. All the results reported in this work are achieved with QDs in a single layer on the silicon device surface. Figure 2a shows an SEM image of a very sparse density of dots («50 per µm$^2$) positioned on a heterostructure cavity, while Figure 2b shows the intensity of the calculated cavity field-mode at the device surface. Figure 2c shows the microphotoluminescence spectroscopy of QDs at this density coupled to our high-$Q$ photonic crystal cavities at room-temperature. Even with an upper limit of 10 to 20% QDs coupled to the cavity mode due to polarization and spatial/spectral matching requirements afforded by a room-temperature broadened single QD linewidth, we infer remarkably low coupled-QD numbers ranging between 1 and 10. At this density, QD coupling to the cavity mode occurs in between 10 to 20% of our devices. We examine the spectroscopic characterization for three different hole radii in the heterostructures (design radii $r$ = 130, 124, and 118-nm shown in shades of blue) and confirm the blue shift in the photoluminescence peak as predicted in the blue-shifted cavity resonance with increasing hole radii. The QDs are pumped non-resonantly with either a 632 nm or 980 nm laser with excitation fluences estimated between 10 to 300 kW/cm$^2$, collected with a 100× (numerical aperture of 0.7) objective, and dispersed with a 32-cm monochromator into a liquid nitrogen-cooled germanium detector. The high excitation fluences are related to the small absorption cross-sections offered by these dots [47], random linear polarization of the source, and due to focusing issues associated from the non-resonant pumping scheme. At this low QD density coverage, we note photoluminescence of the background QDs is embedded beneath the noise floor, and only QDs with emissions enhanced through the cavity can be observed. We emphasize that with less than 10 estimated PbS QDs coupled to our nanocavities, the intensity contrast over the noise floor is observed in the range of 10 to 15× from our different samples, with the noise floor raised due to the high pump powers from the 980 nm laser, to about twice the detector dark noise. Additionally, in order to confirm that the radiation from the QD is from coupling to the cavity mode, we passively measure the mode at 1560 nm using a cross-polarizer setup utilizing reflection spectroscopy and measure a cavity $Q$ of around 200. By comparing the results of the QD-characterized and passive cavity measurements, we clearly see the effect of the coarse PL measurement resolution (required to observe the single-dot levels; 4 nm) — the peak intensity of emission should actually be much higher than observed. As an additional confirmation of this effect, we use high density QDs (<1000 per $\mu m^2$) to decorate a cavity mode at higher resolution (1 nm), and observe spectra that closely resemble our cross-polarization data.

The strongest intensity contrast is typically observed at the 124-nm hole radii samples, possibly due to QD availability at that wavelength. Due to the strong correlation between the QD-coupled and passive cavity spectra (which emphasizes TE polarization of the observed mode), we are confident that coupling occurs to the heterostructure cavity mode.

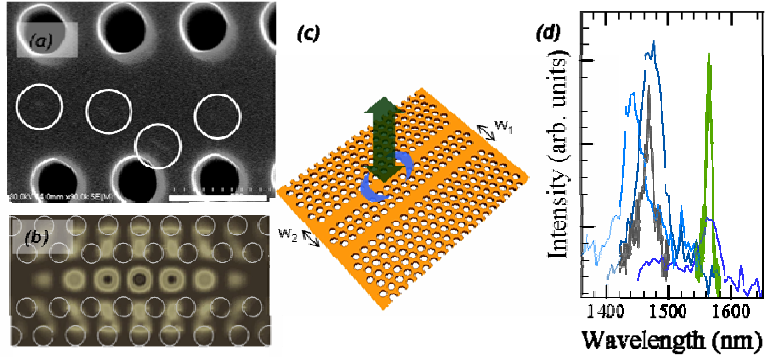

Fig. 2 (a) SEM of less than 50 dots in the heterostructure cavity region. Circles are used to highlight regions of random QD localization after spin-coating. Scale bar: 500 nm (b) $|E_x|^2$ for the calculated field profile (r=124 nm) at silicon slab surface. (c) Schematic of the vertical pump/collection experiment for QD coupling measurements as well as the cross-polarization measurements with the blue region being the region of the confined cavity mode. (d) PL spectra of dots at the cavity region for design radii of 130, 124- 118 nm dots (left to right, blue), at room temperature. The device shown in (a) corresponds to the second mode. Additionally a high resolution scan (1 nm) of the QD photoluminescence (gray) is shown for a large QD density at a cavity region for a device with design radius of 124 nm. The green curve is the passive cross-polarization characterization for the cavity shown at 1560 nm.

We note that AFM topography of the devices confirm that our PbS QD-induced surface roughness over the cavity is less than 4-nm (root-mean-squared), on parity with state-of-the-art InAs cavity-QD systems [48], and confirmed the very few QDs on the devices. We also do not observe any localization of QDs on the hole-sidewalls or on the bottom (Si/SiO$_2$) interface.

## 4. Cryogenic Measurements

We next examine cryogenic tuning of the exciton and cavity lines, ranging from 4K to 300K, as shown in Figure 3. The corresponding typical QD densities (<1000 per $\mu m^2$; still in a monolayer) for these measurements are shown in Figure 4a. The relatively high concentration of QD helps to map out the changes in the background PL spectrum with changing temperature. We confirm that the ensemble QD line red-shifts with decreasing temperature with an overall peak shift from 1460 nm at room-temperature to 1600 nm at 4K. We observe that the shift is piecewise linear as reported in other studies [49]. The red-shift in emission peak is consistent with calculations presented in literature for lead-salt quantum dots with 6 nm diameters [50]. The PL intensity at the peak of QD emission increases as the temperature is decreased from room-temperature until a maximum intensity is observed at 160 K, with a decrease in peak intensity for further reduction in temperature. For solely the exciton state, we also note that the strongest photoluminescence is observed at 160K, possibly due to redistribution of carriers in the presence of defects and thermal excitation [49] although ambient conditions inside the cryostat may also be responsible. Gaussian approximations to the QD emission, and Lorentzian fits to the cavity mode (that underestimate the cavity $Q$ due to the limited spectrometer resolution) are included in Figure 3 for clarity.

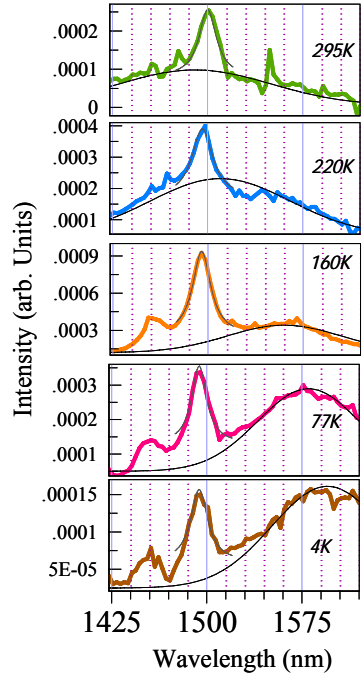

Fig. 3. PL spectra of dots at cavity (r=124 nm) at 4, 77, 160, 220 and 295 K, showing the cavity mode being decorated at all temperatures. Gaussian fits to the QD PL spectrum, as well as Lorentzian fits to the cavity line are also shown.

We find in the QD spectroscopy measurements for this device that the cavity mode remains visible at all scanned temperatures, with intensity enhancements (2 to 4×) over the background QDs as shown in Figure 4b. The observed enhancements are higher at low temperature. The cavity line has a slight blue-shift (60 pm/K) when cooled, and the blue-shift is attributed to the well-known silicon refractive index-dependence of $1.86\times10^{-4}K^{-1}$ [51]. However the results are not typical of every device. At room temperature, the QDs at this density in all devices can couple effectively to the cavity modes through phonon-mediated interactions (room-temperature dephasing). At cryogenic temperatures, QD coupling to the cavity mode is observed to be less efficient with the mode and is only observed in 20% of devices which might arise from a reduced single dot linewidth at 4K, although this effect has not been demonstrated for single PbS QDs [46] in literature. The effectiveness of the cavity in redirecting incoherent QD radiation at room-temperature (where the QD linewidths are broad) to the cavity mode is promising for many quantum information systems applications [52].

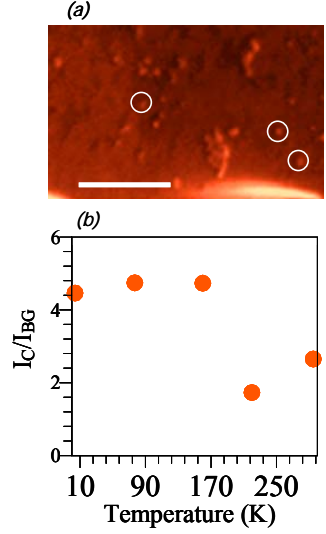

Fig. 4 (a) SEM of typical QD coverage for the results in Figure 3, showing single QDs at concentrations of <1000 per $\mu m^2$. Scale bar: 100 nm (b) QD intensity contrast ($I_c$) over the background ($I_{BG}$) as a function of temperature.

Due to the moderate $Q$ of the devices, the maximum Purcell factors [27],

$$F_P = \frac{3}{4\pi^2}\left(\frac{\lambda}{n}\right)^3\left(\frac{Q}{V}\right) \quad (1)$$

are calculated to be between 12 and 20. The Purcell factor expression is ideally suited for a simple two-level system, and it remains to be verified whether the complex exciton structure of PbS QDs [45, 53, 54] presents a different domain for cavity QED. The high visibility (10-15×) of the very low density of QDs far above the uncoupled QDs is not purely through enhanced collection efficiencies for the cavity-emitted photons (<8%), and strongly suggest spontaneous emission enhancements.

## 5. Saturation Spectroscopy

In order to further support this spontaneous emission enhancement,, a delayed onset of emission saturation for QDs coupled to the cavity mode can be used to demonstrate the Purcell effect [48, 55] based on the idea that a QD coupled well with the cavity mode exhibits a faster radiative recombination rate through the Purcell effect. On resonance with the cavity, mode, it should therefore take more photons to saturate the QD (ground state) emission. This approach was chosen instead of a direct QD lifetime measurement due to the low photon counts and higher dark counts in the near-infrared. We perform saturation spectroscopy of few QDs coupled to the cavity at 1.55-um wavelengths at 77K, comparing between dots at the resonance peak and away from the resonance peak as the only experimental possibilities. Figure 5a shows the sample used in this study, with measured QD coverage of approximately 50 dots per $\mu m^2$ derived from the SEM. Figure 5b shows the PL spectrum of dots, showing a peak at the cavity resonance at 1505 nm with an intensity contrast of 15 at 77K. The estimated number of QDs with emission in the observed spectrum in this experiment is between 5 and 10. The power-saturation of QDs at the resonance peak (1505 nm) and off-peak (1515 nm) is shown in Figure 5c. The photoluminescence intensity at 1515 nm shows a linear increase with pump power at low excitation, and then a saturation of the signal at high pumping rate.

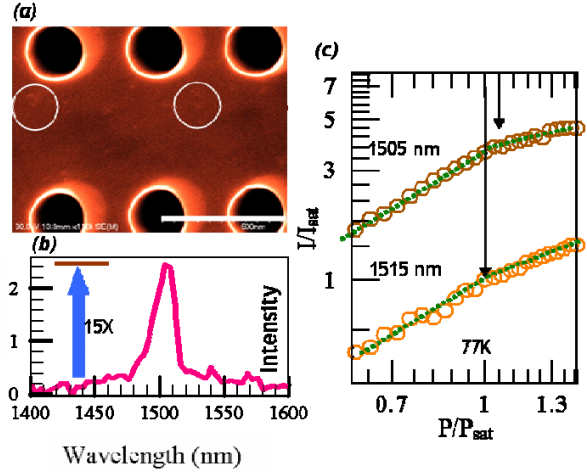

Fig. 5 (a) SEM of approximately 50 QD per μm² at the cavity mode (scale bar: 500 nm) for the device used for power-saturation measurements (different from device shown in Figure 3). (b) PL spectra of QDs at the cavity showing the cavity mode with an intensity contrast of >15 over the background. (c) Power saturation measurements for dots at 1505 and 1515 nm showing a delayed onset of saturation (shown by the arrows) for dots at the peak of emission. This is performed at 77K. The value of $P_{sat}$ is estimated around 5 mW.

We use $P_{sat}$ to denote the power at which saturation occurs for these dots that are slightly detuned from the peak of the cavity mode, meaning that the rate of exciton occupation exceeds the rate of radiative recombination in these dots. We then examine the signal nearly exactly on-resonance with the cavity mode and observe a slightly delayed onset of saturation where a ~10% increase in pumping rate ($P_{on} / P_{off}$) is observed as shown by arrows in Figure 4c. A comparison with QDs completely off-resonance with the cavity line would show further differences in saturation, but due to the very low (less than 50 QD per μm²) dot densities, the photon counts are below the dark counts in the near-infrared. We note that, since the QDs are off-resonantly pumped, the pumping rate is matched for both cases. Moreover, to ascertain the excited saturation against other artifacts such as QD photobleaching, measurement drift or nonlinear absorption, we also vary the pump powers non-monotonically in the measurements. For even lower QD densities such as at a single QD per μm², the delayed onset of saturation can offer an opportunity to experimentally observe QD radiation above the noise floor using high excitation powers. While the results do not allow for a direct estimation of the Purcell factor, the observed delayed saturation onset (10%) for the dots emitting at the peak of the cavity emission qualitatively shows a spontaneous emission enhancement on resonance with the cavity mode.

We note that we did not observe blinking for our few QDs. Although blinking of colloidal QDs can be perceived as a drawback for single photon source applications, recent remarkable efforts have significantly suppressed blinking in nanocrystal QDs, through growth of a thick shell or modification of surface environment [56, 57]. The several coupled QDs estimated here are based on SEM imaging, and it is further likely that some QDs are not active. The devices are therefore expected to be already approaching single dot operation. Compared to smaller PbS QDs with lower photostability, we find these dots to be highly stable under laser excitation, with similar emission levels from the cavity region. We emphasize that the ability to isolate a few quantum dots at the cavity region based on solvent-dilution and post-fabrication integration of QD is an extremely powerful technique. For a single-QD device, the ability to exchange dots based on whether the dot and cavity emission match is not possible in self-assembled semiconductor systems, but is possible here, through selective e-beam lithography techniques that we have proposed earlier. Q values in excess of 1000 are easily achievable in the silicon photonic crystal cavity system, and may allow larger photon counts for time-resolved lifetime measurements for a sparse strongly-enhanced sample through an enhanced Purcell effect, and for photon coincidence measurements of the single Pb-chalcogenide quantum dot at 1.55-um, and for further elucidation of the radiative dynamics of single quantum dots.

In summary, we present our observations of few (1 to 10) strongly quantum-confined PbS-chalcogenide QDs coupled to high-$Q$ heterostructured photonic crystal cavities with efficient performance at room-temperature, presenting a clear improvement from previous studies with ensemble QDs. The examined QD-cavity system exhibits markedly enhanced emission (at 4K and higher temperatures), as evidenced by the sharp intensity enhancements on-resonance for a few QDs, and excited state saturation on-resonance with the mesoscopic cavity mode. These interactions permit the advancement of hybrid silicon-based mesoscopic systems in the context of quantum information processing.

## 6. Acknowledgments

The authors acknowledge helpful discussions with F. Sun, K. Srinivasan, M. Rakher, S. Jockusch, N. Turro, and R. L. Williams, and lithography fabrication from M. Yu and D.-L. Kwong at the Institute of Microelectronics in Singapore. The authors acknowledge funding support from the NSF CAREER program, DARPA MTO, and the New York State Office of Science, Technology and Academic Research. C. W. W. is supported as part of the Energy Frontier Research Center funded by the U.S. Department of Energy, Office of Science, Office of Basic Energy Sciences under Award Number DE-SC0001085.